\documentclass[preprint,aps,superscriptaddress,nofootinbib]{revtex4-1}
\usepackage{graphicx}
\usepackage[toc,page]{appendix}
\usepackage{braket}
\newcommand{\del}{\partial}

\begin{document}
\title{Degenerate spacetimes in first order gravity}

\author{Romesh K. Kaul}
\email{kaul@imsc.res.in}
\affiliation{The Institute of Mathematical Sciences, Chennai-600113, INDIA}

\author{Sandipan Sengupta}
\email{ssandipan@iitgn.ac.in}
\affiliation{Indian Institute of Technology, Gandhinagar-382355, INDIA}

\begin{abstract}
We present a systematic framework to obtain the most general solutions of 
the equations of motion in first order gravity theory with degenerate 
tetrads. There are many possible solutions. Generically, these exhibit 
non-vanishing torsion even in the absence of any matter coupling. 
These solutions are shown to contain a special set of eight configurations 
which are associated with the homogeneous model three-geometries of 
Thurston.

\end{abstract}
  
\maketitle

\section{Introduction}
The usual theory of gravity based on Einstein-Hilbert action functional 
involves invertible metric. Solutions of the vacuum equation of motion are, 
by construction, torsion-free. On the other hand, first-order gravity  
based on Hilbert-Palatini action accommodates invertible as well as 
non-invertible tetrad configurations. The phase containing 
degenerate tetrads can support solutions of the vacuum equations 
of motion with torsion.  In the quantum theory in first order 
formalism, configurations with both invertible and non-invertible 
tetrads are to be integrated over in the functional integral.

Gravity theory with degenerate metrics has evoked interest for 
a long time \cite{einstein, hawking,regge,henneaux,  
tseytlin, ashtekar,  bengtsson, madhavan,jacobson,romano, 
jacob,horowitz}. These metrics  are expected to be 
relevant to the discussion of topology change 
\cite{hawking, tseytlin, ashtekar,  horowitz, wheeler, geroch}. 
Such a topology change may 
have a quantum and even a classical origin \cite{horowitz}.

In this article, we shall present, in the first order formalism, 
a detailed analysis of degenerate tetrads with one zero eigenvalue. 
An elaborate procedure to solve the equations of motion will be
developed. In particular, a set of eight explicit solutions of 
the equations of motion of pure gravity will be presented. These 
are associated with eight independent homogeneous model 
three-geometries of Thurston \cite{thurston, scott, molnar}. 
These include, besides the three isotropic constant curvature 
three-geometries $E^{}_3$, $S^{}_3$ and $H^{}_3$, others which 
are homogeneous but not isotropic. It is remarkable that 
all such degenerate solutions of four dimensional gravity theory 
are not generically torsion-free.

Examples of degenerate tetrad configurations as solutions of 
equations of motion have appeared earlier in the interesting 
work of Tseytlin \cite{tseytlin}. In particular, two explicit 
solutions reported in this reference correspond to two special cases, 
$S^{}_3$ and $S^{}_2 \times R$, as discussed in Section V.

The article is organised as follows. In Section II, we recall 
Hilbert-Palatini action functional without any 
cosmological constant or matter fields and write down the 
consequent equations of motion. Section III outlines the standard 
analysis for invertible tetrads 
to demonstrate the well known fact that such a theory is  
equivalent to the usual theory based on Einstein-Hilbert action. 
The equations of motion are exactly   same as the vacuum 
Einstein field equations. Section IV contains an elaborate discussion 
of degenerate tetrads with one zero eigenvalue. Equations of motion 
are shown to exhibit many possible solutions. Eight explicit 
solutions corresponding to Thurston's homogeneous three-geometries 
are displayed in detail in Section V. Next, the nature of 
the underlying geometry of these degenerate solutions is argued to 
be represented by Sen Gupta geometry \cite{sengupta} in Section 
VI. Finally, some concluding remarks are presented in Section VII. 
An appendix contains details of the calculations used earlier 
in Section IV.   
\section{Hilbert-Palatini action}
Euclidean gravity in the first order formulation is described in 
terms of tetrad fields $e_\mu^I$ and connection fields
$\omega_\mu^{~IJ}$ corresponding to the local Lorentz 
group SO(4). Both these sets of fields are treated as independent 
in the Hilbert-Palatini action functional:
\begin{eqnarray}\label{HP}
S~=~\frac{1}{8\kappa^2}~\int d^4
 x~\epsilon^{\mu\nu\alpha\beta}_{}\epsilon_{IJKL}^{}
e_{\mu}^I e_\nu^J R_{\alpha \beta}^{~~~KL}(\omega)
\end{eqnarray}
where the curvature $R_{\mu \nu}^{~~~IJ} (\omega)=\del_{[\mu}
\omega^{~IJ}_{\nu]}+\omega^{~IK}_{[\mu}\omega^{~KJ}_{\nu]}$ 
is the field strength of the gauge connection
$\omega_{\mu}^{~IJ}$ of the local SO(4) symmetry of Euclidean 
gravity. Here the Greek indices $\mu\equiv(a,\tau)$ are associated 
with the spatial coordinates $a\equiv(x,y,z)$ and Euclidean time
coordinate $\tau$. Internal indices are $I\equiv(i,4),~i=1,2,3$.
Completely antisymmetric epsilon symbols take constant values $0$ and 
$\pm 1$ with $\epsilon^{xyz\tau}_{}=+1$ and $\epsilon_{1234}^{}=+1$. 
Internal indices are raised and lowered by the flat metric
$\eta^{IJ}_{}=\delta^{IJ}_{}=\eta_{IJ}^{}$.

Euler-Lagrange equations of motion are obtained by varying the 
action (\ref{HP}) with respect to $\omega_\mu^{~IJ}$   and $e_\mu^I$  
independently:
\begin{eqnarray}
\frac{\delta S}{\delta\omega_{\beta}^{~IJ}}~&:&~~~~~~\epsilon^{\mu\nu\alpha\beta}_{}
\epsilon_{IJKL}^{} e_\mu^K D^{}_{\nu}(\omega)e_{\alpha}^L~=~0\label{eomi}\\
\frac{\delta S}{\delta e_{\mu}^{I}}~&:&~~~~~~
\epsilon^{\mu\nu\alpha\beta}_{}\epsilon_{IJKL}^{}
e_\nu^J R_{\alpha\beta}^{~~~KL}(\omega)~=~0\label{eomii}
\end{eqnarray}
An equivalent way to display these equations of motion is:
\begin{eqnarray}\label{eom1}
e_{[\mu}^{[I} D^{}_{\nu}(\omega)e_{\alpha]}^{J]}~&=&~0\\
e_{[\mu}^{[I} R_{\nu\alpha]}^{~~~JK]}(\omega)~&=&~0\label{eom2}
\end{eqnarray}
We need to solve these equations for the tetrads and connections. Since the 
Hilbert-Palatini action functional (\ref{HP}) accommodates both invertible 
and non-invertible tetrads, we may consider these two cases separately.

\section{Invertible tetrads}
For tetrads with $\det e_\mu^I\neq 0$, inverse tetrad $e^\mu_I$ is given by
$e^\mu_I e_\nu^I=\delta_\nu^\mu$, $e^\mu_I e_\mu^J=\delta^J_I$. 
Multiplying eq.(\ref{eom1}) by inverse tetrads, it is straightforward to 
check the following identities:
\begin{eqnarray*}
e^\mu_I e_{[\mu}^{[I}D^{}_{\nu} (\omega)e_{\alpha]}^{J]} ~&\equiv &~
D^{}_{[\nu} (\omega) e^J_{\alpha ]} - e^J_\nu \left( e^\mu_I D^{}_{[\alpha}
(\omega) e^I_{\mu]}\right)+ e^J_\alpha \left( e^\mu_I D^{}_{[\nu} 
e^I_{\mu]}\right) ~=~0\\
\mathrm{~and~}~~~~~ e^\mu_I e^\nu_J e_{[\mu}^{[I}D^{}_{\nu}
(\omega)e_{\alpha]}^{J]} ~&\equiv &~ - 4 ~ e^\mu_I D^{}_{[\alpha}
(\omega)e^I_{\mu]}~=~0
\end{eqnarray*}
From these, it readily follows that, for invertible tetrads, 24 equations of 
motion in (\ref{eom1}) are equivalent to the fact that torsion is zero:
\begin{eqnarray}
T_{\mu\nu}^I~\equiv ~D^{}_{[\mu}(\omega) e_{\nu]}^I~=~0
\end{eqnarray}
As is well known, these 24 equations can in turn be solved for 24 
connection fields showing that these are not independent but can be written 
in terms of the tetrad fields as:
\begin{eqnarray}\label{omega(e)}
\omega_{\mu}^{IJ}=\omega_{\mu}^{IJ}(e)\equiv \frac{1}{2}\left[e^\nu_I \del^{}_{[\mu}e_{\nu]}^{J}-e^{\nu}_J \del^{}_{[\mu}e_{\nu]}^{I} - 
e_{\mu}^K e^{\lambda}_I e^{\rho}_J \del^{}_{[\lambda}e_{\rho]}^{K}\right]
\end{eqnarray}

 Other set of 16 equations of motion in (\ref{eom2}), by multiplying
  with $e^\mu_I e^\nu_J$, yield the standard 16 equations of motion:
 \begin{eqnarray}
 R_\alpha^{~K}~-~\frac{1}{2}e_\alpha^K R~ =~ 0
 \end{eqnarray}
 where $R_\alpha^{~K}\equiv e^\mu_I R_{\mu\alpha}^{~~~IK}(\omega)$ 
 and $R\equiv e^\alpha_K R_\alpha^{~K}$. These equations are same 
as Einstein field equations. This follows readily by realizing that 
the local Lorentz field strength, for invertible tetrads, is related 
to the Riemann curvature as :
 \begin{eqnarray}
 R_{\mu\nu}^{~~IJ}(\omega)e_{\lambda}^I
  e^{\rho}_J~=~R_{\mu\nu\lambda}^{~~~~\rho}(\Gamma)
 \end{eqnarray}
 
 Thus the first order formalism for invertible tetrads is 
 exactly equivalent to the second order formalism based on 
 Einstein-Hilbert action functional.

 It is important to notice that for invertible tetrads, solutions of
equations of motion would all be torsion-free. 

 As stated earlier, Hilbert-Palatini action functional (\ref{HP})
 and also the equations of motion  (\ref{eomi}, \ref{eomii}) 
 or (\ref{eom1}, \ref{eom2}) are well
 defined both for invertible tetrads ($\det e_\mu^I\neq 0$) and 
non-invertible tetrads ($\det e_\mu^I=0$). Unlike the case above 
where $\det e_\mu^I\neq 0$, any solution of   equations of 
motion with  degenerate tetrads can in general possess torsion.
Degenerate tetrads   can have one or more zero eigenvalues. 
We shall consider the case of tetrads with only one zero 
eigenvalue here.
 
 \section{Degenerate tetrads with one zero eigenvalue}
Through appropriate local $SO(4)$ rotations and general coordinate transformations, any degenerate tetrad $e_\mu^I$ with one zero eigenvalue can be cast as an   invertible $3\times 3$ block of triads $e_a^i$ ($a=x,y,z$ and $i=1,2,3$) with $e_\tau^I=e_a^4=0$ as follows:\begin{eqnarray}\label{tetrad}
e_\mu^I~=~\left(\begin{array}{ccc}
e_a^i & 0\\
0 & 0\end{array}\right) 
\end{eqnarray}
The four-dimensional metric is:
\begin{eqnarray*}
g_{\mu\nu}~=~e_\mu^I e_\nu^I~=~\left(\begin{array}{ccc}
g_{ab} & 0\\
0 & 0 
\end{array}\right),~~~~~g_{ab}=e_a^i e_b^i,~ 
\end{eqnarray*}
We denote the determinant of the triad as $e$,
 $det ~e_a^i\equiv e ~(\neq 0)$ and its inverse as ${\hat e}^a_i$~;
~ ${\hat e}^a_i e^i_b = \delta^a_b, ~ {\hat e}^a_i e^j_a = \delta^i_j.$
Note that the triad fields $e_a^i$ and the inverse ${\hat e}^a_i$ depend 
on all four spacetime coordinates $(x,y,z,\tau)$. The four 
dimensional infinitesimal length element is: $~ds^2_{(4)}
=0+g_{ab} dx^a dx^b$.

Let us analyse the set of 24 equations in (\ref{eom1}) for such
degenerate tetrads. Unlike the case of invertible tetrads where 
these equations can be solved for all the 24 components of the
connection fields $\omega_\mu^{~IJ}$ as in eqn.(\ref{omega(e)}), 
here for the degenerate tetrads (\ref{tetrad}), eqns (\ref{eom1})
cannot be solved for all the components.

As shown in the Appendix, eqns (\ref{eom1}) can be solved to yield 
the following constraints for  triads $e_a^i$ and   connection
fields $\omega_{\mu}^{~IJ}$:
\begin{eqnarray}
&& D_{\tau}(\omega)e_a^k=0 \mathrm{~~~where~~} \label{e1}
\omega_{\tau}^{~ij}=\bar{\omega}_{\tau}^{~ij}(e)\equiv
 {\hat e}^a_i\del^{}_{\tau}e_a^j = e^i_a \partial^{}_\tau {\hat e}^a_j\\
&& \omega_{\tau}^{~4k}=0; ~~~~~~\omega_{a}^{~4k}\equiv M_a^k=M^{kl} e_a^l
 \mathrm{~~with~~}M^{kl}=M^{lk}  \label{e2}\\
 ~\mathrm{and}~~~~~~~~ && \omega_{a}^{~ij}=\bar{\omega}_{a}^{~ij}(e)+\kappa_a^{~ij};~~~\kappa_a^{~ij}\equiv
\epsilon^{ijk}N_a^k=\epsilon^{ijk}N^{kl}e_a^l~
~\mathrm{with~~}N^{kl}=N^{lk}
\nonumber\\
&& \bar{\omega}_a^{~ij}(e)\equiv 
\frac{1}{2}\left[{\hat e}^b_i\del^{}_{[a}e_{b]}^j
-{\hat e}^b_j\del^{}_{[a}e_{b]}^i -  e_a^l{\hat e}^b_i{\hat e}^c_j
\del^{}_{[b}e_{c]}^l\right]\label{omega-bar}
\end{eqnarray}
Here ${\bar \omega}^{~ij}_a(e)$ and $\kappa_a^{~ij}$
are the torsion-free Levi-Civita connection  and  
contortion fields respectively.

These equations state that the triads $e_a^i$ are covariantly 
conserved with respect to $\tau$ and the connection components
$\omega_{\tau}^{~4k}$ are fixed to be zero. Of the 9 independent 
fields $\omega_a^{~4k}$, three represented by the antisymmetric part 
of the matrix $M^{ij}$ are zero and other six represented by the
symmetric matrix $M^{ij}~ ( =M^{ji})$ are not determined at all. 
Similarly, of the 9 components of the contortion fields
$\kappa_a^{~ij}$, six as represented by the symmetric matrix 
$N^{ij}$ are left undetermined. Thus, for degenerate tetrads
(\ref{tetrad}), eqns (\ref{eom1}) fix only 12 independent fields in
 $\omega_\mu^{~IJ}$ and leave other 12, as encoded by two symmetric
matrices $M^{ij}$ and $N^{ij}$, undetermined. Some of these will be 
further fixed by other equations of motion (\ref{eom2}) 
as discussed below.

Notice that eqns (\ref{e1}) imply that the three-metric is $\tau$
independent: $\del_{\tau}g_{ab}\equiv D_{\tau}(\omega)(e_a^i e_b^i)=0$.
Therefore, $\tau$ dependence  of   triads $e_a^i$ is 
only a pure gauge
artifact and can be rotated away by an $SO(3)$ transformation. That 
is, for an appropriate orthogonal matrix $O^{ij}$, it is 
always possible to write:
\begin{eqnarray}
&& e_a^i=O^{ij}{e'}_a^{j},~~~~\bar{\omega}_{\mu}^{~ij}=O^{il}O^{jk}
{\bar{\omega'}}_\mu^{~lk}+O^{il}\del_{\mu}O^{jl}  \nonumber\\
&& \mathrm {such~ that~~~~}\del_{\tau}e_a^{'i}=0,  
~~~~~\del_{\tau}\bar{\omega'}_{a}^{~ij}(e')=0 
~~~~~\mathrm{and~}~~ \bar{\omega'}_{\tau}^{~ij}=0~\label{gauge2}
\end{eqnarray}

As shown in Appendix, the 16 tetrad equations of motion in
(\ref{eom2}), for degenerate tetrads (\ref{tetrad}), are equivalent 
to the following four sets of 3, 9, 3 and 1 equations respectively:
\begin{eqnarray}
{\hat e}^a_i R_{\tau a}^{~~ij}(\omega)~&=&~0 \label{R1}\\
R_{\tau a}^{~~4k}(\omega)~&=~ &D_{\tau}(\omega)M_a^k~=0~ \label{R2}\\
{\hat e}^a_k R_{ab}^{~~4k}(\omega)~&=& ~\left(e^l_b 
{\hat e}^a_i-\delta_b^a 
\delta^l_i\right)D_{a}(\bar{\omega})M^{il}=0\label{R3}\\
{\hat e}^a_i {\hat e}^b_j {\bar R}_{ab}^{~~ij}(\bar{\omega})&+&\left(M^{ij}M^{ji}-M^{ii}M^{jj}\right)+\left(N^{ij}N^{ji}-N^{ii}N^{jj}\right)~= ~
0\label{R4}
\end{eqnarray}
where $D_a(\bar{\omega})M^{il}\equiv \del_{a}M^{il}+\bar{\omega}_a^{~ij}(e)
M^{jl}+\bar{\omega}_a^{~lj}(e) M^{ij}$ and $\bar{R}_{ab}^{~~ij}(\bar{\omega})$ is the curvature for the torsion-free Levi-Civita spin-connection $\bar{\omega}_a^{~ij}(e)$ of (\ref{omega-bar}):
\begin{eqnarray}
\bar{R}_{ab}^{~~ij}(\bar{\omega})~\equiv~ \del^{}_{[a}\bar{\omega}_{b]}^{~ij}+\bar{\omega}_{[a}^{~il}\bar
{\omega}_{b]}^{~lj} \nonumber
\end{eqnarray}

Equations (\ref{R1}) is identically valid for all configurations 
which satisfy eqns (\ref{e1}-\ref{omega-bar}). To show this, note 
that $R_{\tau a}^{~~ij}(\omega)=\bar{R}_{\tau a}^{~~ij}
(\bar{\omega})+D_{\tau}(\bar{\omega})\kappa_{a}^{~ij}$ 
where $\bar{R}_{\tau a}^{~~ij}(\bar{\omega})\equiv
\del_{[\tau}\bar{\omega}_{a]}^{~ij}+\bar{\omega}_{[\tau}^{~il}
\bar{\omega}_{a]}^{~lj}$. We can write $\bar{R}_{\tau a}^{~~ij}
(\bar{\omega})=O^{il}O^{jk}\bar{R'}_{\tau a}^{~~lk}(\bar{\omega'})$ 
where the gauge rotated primed quantities are as defined in eqns.
(\ref{gauge2}). Now since $\bar{\omega'}_{\tau}^{~ij}(e')=0$ and 
$\del_{\tau}\bar{\omega'}_{a}^{~ij}(e')=0$ for the primed connections 
of (\ref{gauge2}), the curvature $\bar{R'}_{\tau a}^{~~ij}
(\bar{\omega'})\equiv 
\del_{[\tau}\bar{\omega'}_{a]}^{~ij}+\bar{\omega'}_{[\tau}^{~il}
\bar{\omega'}_{a]}^{~lj}\equiv0$ and hence $\bar{R}_{\tau a}^{~~ij}
(\bar{\omega})=0$. This thus implies: $R_{\tau a}^{~~ij}(\omega)=D_{\tau}
(\bar{\omega})\kappa_a^{~ij}$. Contracting with ${\hat e}^a_i$, 
we note that ${\hat e}^a_i 
R_{\tau a }^{~~ij}(\omega)={\hat e}^a_i D_{\tau}
(\bar{\omega})\kappa_a^{~ij}=D_{\tau}
(\bar{\omega})\left({\hat e}^a_i \kappa_a^{~ij}\right)=0$ 
because ${\hat e}^a_i \kappa_a^{~ij}=0$ 
for $\kappa_a^{~ij}=\epsilon^{ijk}N^{kl}e_a^l$ where $N^{kl}=N^{lk}$.

Next, using eqns.(\ref{e1}), we note that the 
constraints (\ref{R2}) and (\ref{R3}) are solved by the choice 
\begin{eqnarray}
M_a^i=\lambda e_a^i  ~~~\Rightarrow ~~~M^{ij} \equiv 
M^i_a {\hat e}^a_j =\lambda \delta^{ij}
\end{eqnarray}
where $\lambda$ is a spacetime constant. This further implies that 
\begin{eqnarray}
M^{ij}M^{ji}-M^{ii}M^{jj}=-6\lambda^2
\end{eqnarray}
Using this the last constraint (\ref{R4}) can then be recast as:
\begin{eqnarray}\label{zeta1}
\zeta~=~6\lambda^2-{\hat e}^a_i {\hat e}^b_j\bar{R}_{ab}^{~~ij}
(\bar{\omega})
\end{eqnarray}
where \begin{eqnarray}\label{zeta2}
\zeta~\equiv ~N^{ij}N^{ji}-N^{ii}N^{jj}~=~2\left(\eta_1^2+\eta_2^2
+\eta_3^2-\alpha\beta-\beta \gamma-
\gamma\alpha\right)
\end{eqnarray}
 for the symmetric matrix 
\begin{eqnarray}\label{N}
N^{ij}~=~\left(\begin{array}{ccc}
\alpha & \eta_3 & \eta_2\\
\eta_3 & \beta & \eta_1\\
\eta_2 & \eta_1 & \gamma\end{array}\right) 
\end{eqnarray}

We conclude this Section, by noting that the action (\ref{HP}) 
for any configuration with degenerate tetrads (\ref{tetrad}) 
satisfying the equations of motion   is zero:
\begin{eqnarray}
S~&=&~\frac{1}{8\kappa^2}\int d^4 x~\epsilon_{}^{\mu\nu\alpha\beta}\epsilon^{}_{IJKL}
e_{\mu}^I e_\nu^J R_{\alpha \beta}^{~~KL}(\omega)\nonumber\\
~&=&~ \frac{1}{2\kappa^2}\int d^4 x~\epsilon_{}^{abc}\epsilon^{}_{ijk}
e_a^i e_b^j R_{c\tau}^{~~k4}(\omega)~=~0,
 \end{eqnarray}
where we have used the constraint (\ref{R2}) in the last step.

\section{Explicit solutions with degenerate tetrads}
To obtain explicit solutions of the equations of motion 
(\ref{e1}-\ref{omega-bar}) and (\ref{R1} -\ref{R3}), all we need to do 
is to prescribe a set of triads $e_a^i$ and associated torsion-free 
Levi-Civita spin-connections $\bar{\omega}_a^{~ij}(e)$ and evaluate the 
spatial (three-) curvature scalar ${\hat e}^a_i {\hat e}^b_j 
\bar{R}_{ab}^{~~ij}(\bar{\omega})$ to fix the combination  $\zeta$ of 
eqn.(\ref{zeta1}). There are many possible solutions. A set of 
solutions for homogeneous three-geometries described by the triads can 
be put in eight classes as given by Thurston's model three-geometries 
\cite{thurston}. We shall now display all these eight solutions.

\subsection*{(i) $E_3$ geometry:}
This flat solution is the simplest where, for affine coordinates 
$x^a\equiv (x,y,z)$, the infinitesimal (squared) length element is: 
$ds_{(4)}^2=dx^2+dy^2+dz^2$. The triads here are simply: 
$e_x^1=e_y^2=e_z^3=1$ and all others zero. Corresponding spin-connection
$\bar{\omega}_a^{~ij}(e)=0$ and so is the three-curvature, 
$\bar{R}_{ab}^{~~ij}(\bar{\omega})=0$. The contortion components as given 
by the symmetric matrix $N^{ij}$ are constrained as:
\begin{eqnarray}\label{g1}
\zeta\equiv 2\left(\eta_1^2+\eta_2^2+\eta_3^2-\alpha\beta-\beta\gamma-
\gamma\alpha\right) =6\lambda^2
\end{eqnarray}

\subsection*{(ii) $S_3$ geometry:}
The metric in terms of the angular coordinates $x^a=(\theta,\phi,\chi)$ for 
this spherical three-geometry is:
\begin{eqnarray*}ds_{(4)}^2=l^2\left[d\theta^2+
sin^2\theta (d\phi^2+sin^2\phi d\chi^2)\right]
\end{eqnarray*}
The only non-zero components of triad are:
\begin{eqnarray*}
e_\theta^1=l,~~~e_\phi^2=lsin\theta,~~~e_\chi^3=lsin\theta sin\phi
\end{eqnarray*}
Associated torsion-free spin-connections for this set of triads are:
\begin{eqnarray*}
\bar{\omega}_\phi^{~12}=-cos\theta,~~~\bar{\omega}_\chi^{~23}=-
cos\phi,~~~\bar{\omega}_\chi^{~31}=cos\theta sin\phi
\end{eqnarray*}
and all others zero. This is a constant curvature three-geometry with 
the   curvature components   given by $\bar{R}_{ab}^{~~ij}
(\bar{\omega})=\frac{1}{l^2}e_{[a}^i e_{b]}^j$ so that the spatial 
curvature scalar is ${\hat e}^{a}_i {\hat e}^{b}_j\bar{R}_{ab}^{~~ij}
(\bar{\omega})=\frac{6}{l^2}$. The contortion components are given by:
\begin{eqnarray*}
N_a^1=l(\alpha,\eta_3 sin\theta,\eta_2 sin\theta 
sin\phi),~N_a^2=l(\eta_3,\beta sin\theta,\eta_1 sin\theta sin\phi), 
N_a^3=l(\eta_2,\eta_1 sin\theta,\gamma sin\theta 
sin\phi)\end{eqnarray*} where the six fields $(\alpha,\beta,\gamma,
\eta_1,\eta_2,\eta_3)$ are as in (\ref{N}). The final constraint 
(\ref{zeta1}) takes the form:
\begin{eqnarray}\label{g2}
\zeta=6\lambda^2-\frac{6}{l^2}
\end{eqnarray}

For the special choice, $N_a^i=l\mu e_a^i$, this $S_3$ configuration 
is exactly a gauge rotated version of the first of the two solutions 
obtained by Tseytlin \cite{tseytlin}.

\subsection*{(iii) $H_3$ geometry:}
The metric for this hyperbolic three-geometry is:
\begin{eqnarray*}
ds_{(4)}^2=\frac{l^2}{z^2}(dx^2+dy^2+dz^2),~~~~z>0
\end{eqnarray*}
Only non-zero components of triad are $e_x^1=e_y^2=e_z^3=\frac{l}{z}$
and those of torsion-free connection are $\bar{\omega}_x^{~31}=\frac{1}
{z}=-\bar{\omega}_y^{~23}$. This is again 
a constant curvature three-geometry with 
the   curvature components   as $\bar{R}_{ab}^{~~ij}
(\bar{\omega})=-\frac{1}{l^2}e_{[a}^i e_{b]}^j$ so that the spatial 
curvature scalar becomes ${\hat e}^a_i {\hat e}^b_j \bar{R}_{ab}^{~~ij}
(\bar{\omega})=-\frac{6}{l^2}$. The contortion is given by 
\begin{eqnarray*}
N_a^1=\frac{l}{z}(\alpha,\eta_3,\eta_2),~~~N_a^2=\frac{l}{z}(\eta_3,\beta,
\eta_1),~~~N_a^3=\frac{l}{z}(\eta_2,\eta_1,\gamma)
\end{eqnarray*}
and the constraint (\ref{zeta1}) becomes:
\begin{eqnarray}\label{g3}
\zeta=\frac {6}{l^2}+6\lambda^2
\end{eqnarray}

\subsection*{(iv) $R \times S_2$ geometry:}
 The metric here is: 
\begin{eqnarray*}
ds_{(4)}^2=dx^2+l^2\left(d\theta^2+
sin^2\theta d\phi^2\right)
\end{eqnarray*}
 Nontrivial triad components are 
 $e_x^1=1,~~e_\theta^2=l,~~e_\phi^3=lsin\theta$ and the only non-zero 
 component of the associated spin connection is 
 $\bar{\omega}_\phi^{~23}=-cos\theta$. There is only one non-vanishing 
 curvature component $\bar{R}_{\theta\phi}^{~~23}
 (\bar{\omega})=sin\theta$ so that the spatial three-curvature scalar 
 is ${\hat e}^a_i {\hat e}^b_j \bar{R}_{ab}^{~~ij}
 (\bar{\omega})=\frac{2}{l^2}$. The 
 contortion components are given by:
\begin{eqnarray*}
N_a^1=(\alpha,l\eta_3  ,l\eta_2 sin\theta  ),~~~N_a^2=(\eta_3,l\beta,l\eta_1 sin\theta ), 
~~N_a^3=(\eta_2,l\eta_1,l\gamma sin\theta)
\end{eqnarray*} 
  and the master constraint (\ref{zeta1}) is:
\begin{eqnarray}\label{g4}
\zeta=6\lambda^2-\frac{2}{l^2}
\end{eqnarray}
This solution is a gauge rotated version of the second solution obtained 
earlier by Tseytlin \cite{tseytlin}.
 
 \subsection*{(v) $R \times H^2$ geometry:}
The infinitesimal arc length square is:
 \begin{eqnarray*}
ds_{(4)}^2=dx^2+\frac{l^2}{z^2}(dy^2+dz^2),~~~~~z>0
\end{eqnarray*}
Non-zero components of triad and the corresponding torsion-free
connection are $e_x^1=1,~e_y^2=e_z^3=\frac{l}{z}$
and $\bar{\omega}_y^{~23}=-\frac{1}{z}$. Curvature has only one 
non-zero component,  $\bar{R}_{yz}^{~~23}(\bar{\omega})=-\frac{1}
{z^2}$, leading to the spatial curvature scalar 
${\hat e}^a_i {\hat e}^b_j 
\bar{R}_{ab}^{~~ij}
(\bar{\omega})=-\frac{2}{l^2}$. The contortion is given by 
\begin{eqnarray*}
N_a^1=(\alpha,\frac{l}{z}\eta_3,\frac{l}{z}\eta_2),~~~N_a^2=(\eta_3,
\frac{l}{z}\beta,\frac{l}{z}\eta_1),~~~N_a^3=(\eta_2,\frac{l}{z}\eta_1,
\frac{l}{z}\gamma)
\end{eqnarray*}
Finally we have the constraint:
\begin{eqnarray}\label{g5}
\zeta=\frac{2}{l^2}+6\lambda^2
\end{eqnarray}

\subsection*{(vi) Sol-geometry:}
Here the metric is:
\begin{eqnarray*}
ds_{(4)}^2=e^{\frac{2z}{l}}dx^2+e^{-\frac{2z}{l}}dy^2+dz^2
\end{eqnarray*} with non-zero components of the triads and spin-connection 
fields as: 
\begin{eqnarray*}
e_x^1&=&e^{\frac{z}{l}},~~~e_y^2=e^{-\frac{z}{l}},~~~e_z^3=1\\
\bar{\omega}_{y}^{~23}&=&-\frac{e^{-\frac{z}{l}}}
{l},~~~\bar{\omega}_{x}^{~31}=-\frac{e^{\frac{z}{l}}}{l}~.
\end{eqnarray*}
Non-vanishing curvature components are
\begin{eqnarray*}
\bar{R}_{xy}^{~~12}(\bar{\omega})=\frac{1}{l^2},~~~\bar{R}_{yz}^{~~23}
(\bar{\omega})=-\frac{e^{-\frac{z}{l}}}{l^2},~~~\bar{R}_{zx}^{~~31}
(\bar{\omega})=-\frac{e^{\frac{z}{l}}}{l^2}
\end{eqnarray*}
so that ${\hat e}^a_i {\hat e}^b_j \bar{R}_{ab}^{~~ij}
(\bar{\omega})=-\frac{2}{l^2}$.
 The contortion fields are  
\begin{eqnarray*}
N_a^1=(\alpha e^{\frac{z}{l}},\eta_3 e^{-\frac{z}{l}},
\eta_2),~~~N_a^2=(\eta_3 e^{\frac{z}{l}},\beta e^{-\frac{z}{l}},
\eta_1),~~~N_a^3=(\eta_2 e^{\frac{z}{l}},\eta_1 e^{-\frac{z}{l}},\gamma)
\end{eqnarray*}
With these, the constraint (\ref{zeta1}) becomes:
\begin{eqnarray}\label{g6}
\zeta=\frac{2}{l^2}+6\lambda^2
\end{eqnarray}
  \subsection*{(vii) Nil-geometry:}
 This geometry is characterized by the metric: 
\begin{eqnarray*}
ds_{(4)}^2=dx^2+dy^2+(dz-\frac{x}{l}dy)^2
\end{eqnarray*}
with non-zero triad components as  $~e_x^1=1;~~e_y^2=1,~~e_y^3=-\frac{x}
{l};~~e_z^3=1$ and the nontrivial components of the inverse as 
$~{\hat e}^x_1=1;~~{\hat e}^y_2=1;~~{\hat e}^z_2=\frac{x}{l},
~{\hat e}^z_3=1$. Non-vanishing components of the torsion-free 
spin connection are:
$\bar{\omega}_y^{~12}=-\frac{x}
{2l^2},~~\bar{\omega}_z^{~12}=- \bar{\omega}_x^{~23}=
-\bar{\omega}_y^{~31}=
\frac{1}{2l}$. These lead to $\bar{R}_{xy}^{~~12}(\bar{\omega})
=-\frac{3}{4 l^2},
~\bar{R}_{yz}^{~~23}(\bar{\omega})=\frac{1}{4 l^2}=\bar{R}_{zx}^{~~31}
(\bar{\omega}),~\bar{R}_{xy}^{~~31}(\bar{\omega})=\frac{x}{4 l^3}$ 
as the only non-zero curvature components. Thus, the curvature 
scalar is ${\hat e}^a_i {\hat e}^b_j 
\bar{R}_{ab}^{~~ij}(\bar{\omega})=-\frac{1}
{2l^2}$. The contortion fields are: 
\begin{eqnarray*}
N_a^1=(\alpha,\eta_3-\frac{x}{l}\eta_2,\eta_2),~~N_a^2=(\eta_3,
\beta-\frac{x}{l}\eta_1,\eta_1),~~N_a^3=(\eta_2,\eta_1-\frac{x}
{l}\gamma,\gamma)
\end{eqnarray*} and the constraint (\ref{zeta1}) reads:
\begin{eqnarray}\label{g7}
\zeta=\frac{1}{2l^2}+6\lambda^2
\end{eqnarray}
\subsection*{(viii) $\widetilde{SL_2 R}$-geometry:}
 The metric is given by \cite{molnar} 
\begin{eqnarray*}
ds_{(4)}^2=dr^2+l^2\left[c^2 s^2 d\theta^2+(d\phi+s^2 d\theta)^2\right]
\end{eqnarray*}
where $c\equiv cosh\left(\frac{r}{l}\right)$ and $s\equiv 
sinh\left(\frac{r}{l}\right)$. The non-vanishing components of triad, 
inverse triad and torsion-free spin connection are:
\begin{eqnarray*}
e_r^1&=&1;~~e_{\theta}^2=lsc,~~e_{\theta}^3=ls^2; ~~ e^3_\phi = l;~\\
 {\hat e}^r_1&=&1;~~{\hat e}^{\theta}_2=\frac{1}{lsc};~~{\hat e}^{\phi}_2=
 -\frac{s}{lc},~~{\hat e}^{\phi  }_3=\frac{1}{l};\\
\bar{\omega}_{\theta}^{~12}&=&-(c^2+2s^2),~~\bar{\omega}_{\phi}^{~12}=-1,~
~\bar{\omega}_r^{~23}=\frac{1}{l},~~\bar{\omega}_{\theta}^{~31}=cs~.
\end{eqnarray*}  
These imply that only the following curvature components are non vanishing:
\begin{eqnarray*}
\bar{R}_{r\theta}^{~~12}(\bar{\omega})=-\frac{7cs}
{l},~~\bar{R}_{\theta\phi}^{~~23}
(\bar{\omega})=cs,~~\bar{R}_{r\theta}^{~~31}(\bar{\omega})=-\frac{s^2}
{l},~~\bar{R}_{\phi r}^{~~31}(\bar{\omega})=\frac{1}{l}\end{eqnarray*}
so that the curvature scalar is ${\hat e}^a_i {\hat e}^b_j 
\bar{R}_{ab}^{~~ij}(\bar{\omega})=-\frac{10}{l^2}$. The contortion 
components are: 
\begin{eqnarray*}
N_a^1=(\alpha,lsc 
\eta_3+ls^2\eta_2,l\eta_2),~N_a^2=(\eta_3,lsc\beta+ls^2\eta_1,l\eta_1),
~N_a^3=(\eta_2,lsc\eta_1+ls^2\gamma,l\gamma)
\end{eqnarray*}
The final constraint (\ref{zeta1}) now is:
\begin{eqnarray}\label{g8}
\zeta=\frac{10}{l^2}+6\lambda^2
\end{eqnarray}

With this we have completed the discussion of various explicit solutions 
associated with Thurston's eight model three-geometries. All these 
solutions generically contain torsion as reflected by the symmetric matrix 
$N^{ij}$ where the contortion is parametrized as 
$\kappa_a^{~ij}=\epsilon^{ijk}N^{kl}e_a^l$. Six component fields of 
symmetric $N^{ij}$ depend on all the four spacetime coordinates ($x,y,z,
\tau$). These are independent   except for one constraint   
  so that the combination $\zeta=(N^{ij}N^{ji}-
N^{ii}N^{jj})$ has fixed values as dictated by the 
condition (\ref{zeta1}) for 
various solutions. For all the eight solutions above, $\zeta$ as given by 
eqns (\ref{g1}-\ref{g8}), is spacetime constant in each case.

To emphasize, unlike the case of invertible tetrads where torsion enter 
into the theory through matter couplings such as fermions, here in the 
phase with degenerate tetrads  torsion is exhibited by 
the solutions even in the case
 of pure gravity without any torsion-inducing matter fields.
 
 Our discussion of degenerate tetrads above has been set up in Euclidean 
 gravity. As is obvious, it holds equally well for Lorentzian signature
 where zero eigenvalue of the tetrad is in the time direction.
 Also, the analysis has a straight forward  generalization 
 even when  the zero  eigenvalue is in a spatial direction where the 
 nontrivial three-geometry would now be  Lorentzian and 
 corresponding changes for 
 three of the six torsional components will appear.  
 
\section{Sen Gupta Geometry}
 The degenerate tetrad solutions we have discussed here do not represent the 
 usual geometry as seen in the Einsteinian gravity. To understand the 
 nature of these solutions, let us go to the flat spacetime limit.
 
 We shall use Lorentzian signature in the discussion that follows. In the 
 flat limit, square of the infinitesimal length element is given by
\begin{eqnarray*}
ds^2_{(4)}=-c^2 dt^2+dx^2+dy^2+dz^2
\end{eqnarray*} 
 Degenerate tetrad with one zero eigenvalue as considered here correspond 
 to the limit where the metric component $g_{tt}\equiv c^2\rightarrow 0$ in 
 this flat spacetime case.
 
 Under a change of frame, the length element stays unaltered:
\begin{eqnarray*}
ds^2_{(4)}=-c^2 dt^2+dx^2+dy^2+dz^2=-c^2 dt^{'2}+dx^{'2}+dy^{'2}+dz^{'2}
\end{eqnarray*}  
 There are two ways of writing transformations which leave $ds^2_{(4)}$ 
 invariant. First is the standard Lorentz transformation:
\begin{eqnarray}\label{LT}
dt'= \frac{dt-\frac{v}{c^2}dx}{\sqrt{1-\frac{v^2}{c^2}}},~~dx'=\frac{dx-
vdt}{\sqrt{1-\frac{v^2}{c^2}}},~~dy'=dy,~~dz'=dz
\end{eqnarray}
 where we have introduced the boost transformation in the $t-x$ plane. Here 
 the parameter $v$,  bounded from above as $v^2<  c^2$,  is the relative 
 velocity between the frames. In other words, $v=\frac{dx}{dt}$ (for 
 $\Delta x'=0$) is the velocity of a fixed point in the primed frame in the 
 spacetime of the unprimed frame. As pointed out by Sen Gupta 
 \cite{sengupta}, there is another transformation which leaves the length 
 element $ds^2_{(4)}$ invariant:
 \begin{eqnarray}\label{ILT}
dt'= \frac{dt-\frac{dx}{w}}{\sqrt{1-\frac{c^2}
{w^2}}},~~dx'=\frac{dx-\frac{c^2}{w}dt}{\sqrt{1-\frac{c^2}
{w^2}}},~~dy'=dy,~~dz'=dz
\end{eqnarray}
 Here the parameter $w$ is bounded from  below as $w^2> c^2$. Despite its 
 dimensions, $w$ is not a relative frame velocity. Since $w=\frac{dx}{dt}$ 
 for $\Delta t'=0$, it rather represents the rate of change of an event 
 that occurs at a fixed time in the primed system as measured in 
 the unprimed  system. The two transformations (\ref{LT}) and (\ref{ILT}) 
 are dual to each other. They go to each other under the changes
 $v\rightarrow \frac{c^2}{w}$ and $w\rightarrow \frac{c^2}{v}$.
 
 The non-relativistic limit of the Lorentz transformation is obtained by 
 taking $c\rightarrow \infty$ limit in (\ref{LT}) to yield the standard 
 Galilean transformation:
\begin{eqnarray}\label{GT}
dt'= dt,~~dx'=dx-vdt,~~dy'=dy,~~dz'=dz
\end{eqnarray}
On the other hand, it is the transformation (\ref{ILT}) that is appropriate
for studying the limit $c\rightarrow 0$. In this limit, as was pointed out 
by Sen Gupta,  transformation (\ref{ILT}) leads to the 
following   dual transformation:
\begin{eqnarray}\label{IGT}
dt'= dt-\frac{dx}{w},~~dx'=dx,~~dy'=dy,~~dz'=dz
\end{eqnarray} 
This transformation \cite{sengupta, levylebond}, 
though analogous to the Galilean 
 transformation (\ref{GT}), yet is different with the roles of space and 
 time interchanged. We may refer to the spacetime with transformation 
 properties (\ref{IGT}) as Sen Gupta spacetime.
 
 The phase of degenerate tetrads in the first order formalism discussed 
 in this article describes the curved spacetime generalizations of the Sen 
 Gupta spacetime. This is in contrast to the phase with invertible tetrads 
 which corresponds to the usual Einstein curved spacetime.
 
\section{Concluding remarks}
The phase containing invertible tetrads in the first order gravity based on 
Hilbert-Palatini action is exactly same as the usual Einstein geometry 
described by the second order formalism based on Einstein-Hilbert action. 
However, in the first order formulation there is another phase containing 
non-invertible tetrads. Thus, even classically the two formalisms are not 
equivalent.

Here we have studied in detail possible degenerate tetrad solutions with 
one zero eigenvalue in first order gravity. Many such solutions are 
possible. A special class of solutions obtained are associated with 
Thurston's eight homogeneous three-geometries. All these solutions 
generically possess torsion without the presence of any matter fields such 
as fermions.

While the solutions with invertible tetrads correspond to the usual 
Einstein geometry, the degenerate ones with one zero eigenvalue are curved 
spacetime generalizations of Sen Gupta (flat) spacetime geometry.

In the quantum theory of gravity we need to integrate over all possible 
configurations, including those with degenerate tetrads, in the functional 
integral as prescribed by Feynman path integral formulation. Such 
non invertible configurations can play an important role in the quantum 
theory. 

  Although our analysis has been presented in the framework of Euclidean 
gravity, it is also valid for Lorentzian gravity where the zero eigenvalue 
of the tetrad is in the time direction. In particular, the eight explicit 
solutions displayed in Section V are valid for this case as well. The 
analysis with the null eigenvalue in a spatial direction is  also  
a mere simple generalization of the analysis elucidated here.

\acknowledgments
R.K.K.   acknowledges the support of Department
of Science and Technology, Government of India, through a J.C.
Bose National Fellowship.
S.S. acknowledges the hospitality and generous support of the Institute of 
Mathematical Sciences, Chennai where part of this work was done. 

\subsection*{{\bf Appendix}}

Here we shall present the details of derivations of eqns.
(\ref{e1}-\ref{omega-bar}) and (\ref{R1}-\ref{R4}).

For the degenerate tetrads (\ref{tetrad}), we may break the 24 equations in 
(\ref{eom1}) into two sets of 18 and 6 equations   respectively as:
\begin{eqnarray}
e_{[\tau}^{[I} D^{}_{a}(\omega)e_{b]}^{J]}&=&0\label{E1}\\
e_{[a}^{[I} D^{}_{b}(\omega)e_{c]}^{J]}&=&0\label{E2}
\end{eqnarray}

It is straightforward to see that eqns.(\ref{E1}) can be recast as 
equivalent 18 equations:
\begin{eqnarray}
D_{\tau}(\omega)\left(e_{a}^{[I}e_{b}^{J]}\right)&=&0\label{E3}
\end{eqnarray}
Taking $I=i$ and $J=4$, these result in nine equations:
$e_{[a}^{i}e_{b]}^{k}\omega_{\tau}^{~4k}=0$. These in turn imply  vanishing 
of $\omega_{\tau}^{~4i}$ as claimed in (\ref{e2}). Again, for $I=i, ~J=j$, 
eqns.(\ref{E3}) lead to the 9 equations 
$D^{}_{\tau}(\omega)\left(e_{a}^{[i}e_{b}^{j]}\right)=0$. 
These are equivalent 
to nine equations $D_{\tau}(\omega)e_{a}^{i}=0$ as claimed in (\ref{e1}). 
These further imply $D^{}_{\tau}(\omega){\hat e}^{a}_{i}=0$ 
and $\del^{}_\tau e=0$ where 
$e=det~ e_a^i$. These equations can be solved for the connection 
components  $\omega_\tau^{~ij}$ as:
\begin{eqnarray}
\omega_\tau^{~ij}=\bar{\omega}_\tau^{~ij}(e)\equiv 
{\hat e}^a_i\del^{}_\tau 
e_a^j=e_a^i\del^{}_\tau {\hat e}^a_j=-{\hat e}^a_j\del^{}_\tau 
e_a^i=- e^j_a\del^{}_\tau {\hat e}_i^a
\end{eqnarray} 

Next, we take $I=i,~J=4$ in eqns.(\ref{E2}) and multiply by 
${\hat e}^a_i$ to show 
that $D^{}_{[b}(\omega)e_{c]}^4\equiv 
\omega_{[b}^{~4k} e_{c]}^k=0$ which in 
turn imply that $\omega _b^{~4k}\equiv M_b^k=M^{kl}e_b^l$ is such that the  
$3\times 3$ matrix $M^{kl}$ is symmetric. Further for $I=i, ~J=j$ in 
(\ref{E2}), multiplying by ${\hat e}^a_i {\hat e}^b_j$, 
it can readily be shown to lead to three conditions:
\begin{eqnarray}\label{E4}
{\hat e}^a_i D^{}_{[c}(\omega)e_{a]}^{i}=0
\end{eqnarray} 
Now let us split the connection fields as 
$\omega_a^{~ij}=\bar{\omega}_a^{~ij}(e)+\kappa_a^{~ij}$ 
where  $\kappa_a^{~ij}\equiv 
\epsilon^{ijk}N_a^k\equiv \epsilon^{ijk}N^{kl}e_a^l$
are the contortion fields and $\bar{\omega}_a^{~ij}(e)$,
as given in eq.(\ref{omega-bar}),  are the torsion-free 
Levi-Civita spin connections   for the triads $e_a^i$:
\begin{eqnarray}
D^{}_{[a}(\bar{\omega})e_{b]}^{i}=0
\end{eqnarray} 
With this,  eqn.(\ref{E4}) can be shown to imply 
that ${\hat e}^a_i \kappa_a^{~ij}
  =0$. This further leads to the fact that   the $3\times 3$ contortion 
 matrix $N^{ij}$ does not have any anti-symmetric part, that is, 
 $N^{ij}=N^{ji}$.

Next, we split the 16 equations in (\ref{eom2}) into two sets of 12 and 4 
equations respectively as:
\begin{eqnarray}
e_{[\tau}^{[I} R_{ab]}^{~~JK]}(\omega)&=&0\label{E5}\\
e_{[a}^{[I} R_{bc]}^{~~JK]}(\omega)&=&0\label{E6}
\end{eqnarray}

In eqns.(\ref{E5}) we take $I=4,~J=j, ~K=k$ and multiply by inverse triads
 to note that ${\hat e}^a_j e_{[\tau}^{[4} R_{ab]}^{~~~jk]}
 (\omega)=4\left[R_{b\tau}^{~~k4}(\omega)+e_b^k 
 \left({\hat e}^a_j R_{\tau a}^{~~4j}
 (\omega)\right)\right]=0$ and 
 ${\hat e}^b_k {\hat e}^a_j e^{[4}_{[\tau} R_{ab]}^{~~jk]}
 (\omega)=16 {\hat e}^a_k R_{a\tau}^{~~4k}(\omega)=0$. 
 This leads to the 9 conditions 
 $R_{a\tau}^{~~4k}(\omega)=0$. Further using the fact that 
 $\omega_{\tau}^{~4k}=0$ as argued above, we note that $R_{a\tau}^{~~4k}
 (\omega)\equiv  -
 \left(\del_{\tau}\omega_a^{~4k}+\omega_{\tau}^{~kl}
 \omega_a^{~4l}\right)\equiv 
 -\left(\del_\tau M_a^k+\omega_\tau^{~kl}M_a^l\right)\equiv -
 D_\tau(\omega)M_a^k$. Thus we have the nine constraints:
 \begin{eqnarray}
 R_{a\tau}^{~~4k}(\omega)=-D_\tau (\omega) M_a^k=0
 \end{eqnarray}

 Again in eqn.(\ref{E5}), we take $I=i,~J=j,~K=k$ and use the fact that 
 ${\hat e}^a_i {\hat e}^b_j e_{[\tau}^{[i}R_{ab]}^{~~~jk]}(\omega)
 =8{\hat e}^a_i R_{\tau a}^{~~ki}(\omega)$ leading us to three constraints:
 \begin{eqnarray}
{\hat e}^a_i R_{\tau a}^{~~ki}(\omega)=0
\end{eqnarray}  

Next, let us take $I=4,~J=j,~K=k$ in eqn.(\ref{E6}) and notice that
${\hat e}^a_j {\hat e}^b_k e^{[4}_{[a} R^{~~~jk]}_{bc]} (\omega) = 8 {\hat 
e}^a_j R^{~~~4j}_{ca} (\omega)$ leading us to three conditions: 
\begin{eqnarray}
{\hat e}^a_k R_{ab}^{~~4k}(\omega)= \left(e_b^l {\hat e}^a_i 
-\delta_b^a \delta_i^l \right)D_a(\bar{\omega})M^{il}=0
\end{eqnarray}
where for the first step we have used $R_{ab}^{~~4k}(\omega)
=D^{}_{[a}(\omega)M_{b]}^k =D^{}_{[a}
(\bar{\omega})M_{b]}^k+\kappa_{[a}^{kl}M_{b]}^l$ and 
${\hat e}^a_k \kappa_a^{kl}=0$ and $M^{il}\equiv 
M_a^i {\hat e}_l^a = M^{li}_{}$.

Finally taking $I=i,~J=j,~K=k$ in (\ref{E6}) and using 
$\epsilon^{abc}\epsilon_{ijk}e_a^iR_{bc}^{~~jk}
=2e {\hat e}^b_j{\hat e}^c_k R_{bc}^{~~jk}$ 
we obtain the last condition as:
\begin{eqnarray}
{\hat e}^a_i {\hat e}^b_j R_{ab}^{~~ij}(\omega)=0\label{E9}
\end{eqnarray}
Expanding $\omega_a^{~ij}=\bar{\omega}_a^{~ij}
(e)+\epsilon^{ijk}N^{kl}e_a^l$, we find that:
\begin{eqnarray} 
R_{ab}^{~~ij}(\omega)=\bar{R}_{ab}^{~~ij}
(\bar{\omega})-\epsilon^{ijk}e^l_{[a}
D^{}_{b]}(\bar{\omega})N^{kl}-\left(M^{il}M^{jk}+N^{il}N^{jk}\right)
e_{[a}^{l}e_{b]}^{k}
\end{eqnarray}
where $\bar{R}_{ab}^{~~ij}
(\bar{\omega})=\del^{}_{[a}\bar{\omega}_{b]}^{~ij}
+\bar{\omega}_{[a}^{~il}\bar{\omega}_{b]}^{~lj}$.  Using this, the 
constraint (\ref{E9}) can be recast as:
\begin{eqnarray}
{\hat e}^a_i {\hat e}^b_j \bar{R}_{ab}^{~~ij}
(\bar{\omega})+\left(M^{ij}M^{ji}-M^{ii} M^{jj}\right)
+\left(N^{ij}N^{ji}-N^{ii} N^{jj}\right)=0
\end{eqnarray}
where we have used the fact that matrix $N^{ij}$ is symmetric.


\begin{thebibliography}{99}

\bibitem{einstein} A. Einstein and N. Rosen, Phys. Rev. {\bf 48} (1935) 73.

\bibitem{hawking} S. Hawking, Nucl. Phys. {\bf B 144} (1978) 349.

\bibitem{henneaux} M. Henneaux, Bull. Soc. Math. Belg. {\bf 31} (1979) 47;\\
M. Henneaux, M. Pilati, C. Teitelboim, Phys. Lett. {\bf B110} (1982) 123;\\
M. Pilati, Phys. Rev. {\bf D26} (1982) 2645;\\
M. Pilati, Phys. Rev. {\bf D28} (1983) 729.

\bibitem{regge} R. D'Auria and T. Regge, Nucl. Phys. {\bf B195} (1982) 308-324. 

\bibitem{tseytlin} A.A. Tseytlin, J. Phys. A: Math. Gen. {\bf 15} (1982) 
L105.

\bibitem{ashtekar} A. Ashtekar, Phys. Rev. {\bf D36} (1987) 1587.

\bibitem{bengtsson} I. Bengtsson, Int. J. Mod. Phys. {\bf A4} 
(1989) 5527;\\
S. Koshti and N. Dadhich, Class. Quantum Grav. {\bf 6} (1989) L223;\\
I. Bengtsson, Class Quantum Grav. {\bf 7} (1990) 27;\\
I. Bengtsson, Class. Quantum Grav. {\bf 8} (1991) 1847.

\bibitem{madhavan} M. Varadarajan, Class. Quantum Grav. {\bf 8} (1991) 11, 
L235.

\bibitem{jacobson} I. Bengtsson, T. Jacobson, Class. Quantum Grav. {\bf 14} 
(1997) 3109; \\
Erratum-ibid. {\bf 15} (1998) 3941.

\bibitem{romano} T. Jacobson, J.D. Romano, Class. Quantum Grav. {\bf 9} 
(1992) L119, gr-qc/9207005;\\
J.D. Romano, Phys.Rev. {\bf D48} (1993) 5676, gr-qc/9306034;\\
M.P. Reisenberger, Nucl. Phys. {\bf B457} (1995) 643, gr-qc/9505044;\\
G. Yoneda, H. Shinkai, A. Nakamichi, Phys. Rev. {\bf D56} (1997) 2086.  

\bibitem{jacob} T. Jacobson, Class. Quantum Grav. {\bf 13} (1996) L111-L116 and
Erratum ibid {\bf 13} (1996) 3269;\\
S.H.S. Alexander and G. Calcagni, Found. Phys. {\bf 38} (2008) 1148;\\
S.H.S. Alexander and G. Calcagni, Phys. Lett. {\bf B 672} (2009) 386.

\bibitem{horowitz} G.T. Horowitz, Class. Quantum Grav. {\bf 8} (1991) 587.

\bibitem{wheeler} J. A. Wheeler, Annals Phys. {\bf 2} (1957) 604;\\ 
J.A. Wheeler, {\it Geometrodynamics}, Academic Press, New York, (1962).

\bibitem{geroch} R.P. Geroch, J. Math. Phys. {\bf 8} (1967) 782. 

\bibitem{thurston} W.P. Thurston, Bull. A.M.S. {\bf 6} (1982) 357;\\
W.P. Thurston, {\it The geometry and topology of 3-manifolds}, Princeton 
University Lecture Notes (1982).

 \bibitem{scott} P. Scott, Bull. London Math. Soc. {\bf 15} (1983) 401.

\bibitem{molnar} E. Molnar, Beitrage zur Algebra und Geometrie {\bf 38} 
No.2 (1997) 261.

\bibitem{sengupta} N.D. Sen Gupta, Nuovo Cimento {\bf 44} (1966) 512.

\bibitem{levylebond} J-M. Levy-Leblond, Ann. Inst. Henri Poincare, 
{\bf 3} (1965) 1. 

\end{thebibliography}
\end{document}